\documentclass{article}
\usepackage{amsmath, amssymb, amsfonts}
\title{Boundary Sources in the Doran - Lobo - Crawford Spacetime}
\author{Hristu Culetu, \\Ovidius University, Dept.of Physics, \\B-dul Mamaia 124, 8700 Constanta, Romania, \\e-mail : hculetu@yahoo.com}

\begin{document}
\numberwithin{equation}{section}
\pagenumbering{arabic}
\maketitle
\newcommand{\fv}{\boldsymbol{f}}
\newcommand{\tv}{\boldsymbol{t}}
\newcommand{\gv}{\boldsymbol{g}}
\newcommand{\OV}{\boldsymbol{O}}
\newcommand{\wv}{\boldsymbol{w}}
\newcommand{\WV}{\boldsymbol{W}}
\newcommand{\NV}{\boldsymbol{N}}
\newcommand{\hv}{\boldsymbol{h}}
\newcommand{\yv}{\boldsymbol{y}}
\newcommand{\RE}{\textrm{Re}}
\newcommand{\IM}{\textrm{Im}}
\newcommand{\rot}{\textrm{rot}}
\newcommand{\dv}{\boldsymbol{d}}
\newcommand{\grad}{\textrm{grad}}
\newcommand{\Tr}{\textrm{Tr}}
\newcommand{\ua}{\uparrow}
\newcommand{\da}{\downarrow}
\newcommand{\ct}{\textrm{const}}
\newcommand{\xv}{\boldsymbol{x}}
\newcommand{\mv}{\boldsymbol{m}}
\newcommand{\rv}{\boldsymbol{r}}
\newcommand{\kv}{\boldsymbol{k}}
\newcommand{\VE}{\boldsymbol{V}}
\newcommand{\sv}{\boldsymbol{s}}
\newcommand{\RV}{\boldsymbol{R}}
\newcommand{\pv}{\boldsymbol{p}}
\newcommand{\PV}{\boldsymbol{P}}
\newcommand{\EV}{\boldsymbol{E}}
\newcommand{\DV}{\boldsymbol{D}}
\newcommand{\BV}{\boldsymbol{B}}
\newcommand{\HV}{\boldsymbol{H}}
\newcommand{\MV}{\boldsymbol{M}}
\newcommand{\be}{\begin{equation}}
\newcommand{\ee}{\end{equation}}
\newcommand{\ba}{\begin{eqnarray}}
\newcommand{\ea}{\end{eqnarray}}
\newcommand{\bq}{\begin{eqnarray*}}
\newcommand{\eq}{\end{eqnarray*}}
\newcommand{\pa}{\partial}
\newcommand{\f}{\frac}
\newcommand{\FV}{\boldsymbol{F}}
\newcommand{\ve}{\boldsymbol{v}}
\newcommand{\AV}{\boldsymbol{A}}
\newcommand{\jv}{\boldsymbol{j}}
\newcommand{\LV}{\boldsymbol{L}}
\newcommand{\SV}{\boldsymbol{S}}
\newcommand{\av}{\boldsymbol{a}}
\newcommand{\qv}{\boldsymbol{q}}
\newcommand{\QV}{\boldsymbol{Q}}
\newcommand{\ev}{\boldsymbol{e}}
\newcommand{\uv}{\boldsymbol{u}}
\newcommand{\KV}{\boldsymbol{K}}
\newcommand{\ro}{\boldsymbol{\rho}}
\newcommand{\si}{\boldsymbol{\sigma}}
\newcommand{\thv}{\boldsymbol{\theta}}
\newcommand{\bv}{\boldsymbol{b}}
\newcommand{\JV}{\boldsymbol{J}}
\newcommand{\nv}{\boldsymbol{n}}
\newcommand{\lv}{\boldsymbol{l}}
\newcommand{\om}{\boldsymbol{\omega}}
\newcommand{\Om}{\boldsymbol{\Omega}}
\newcommand{\Piv}{\boldsymbol{\Pi}}
\newcommand{\UV}{\boldsymbol{U}}
\newcommand{\iv}{\boldsymbol{i}}
\newcommand{\nuv}{\boldsymbol{\nu}}
\newcommand{\muv}{\boldsymbol{\mu}}
\newcommand{\lm}{\boldsymbol{\lambda}}
\newcommand{\Lm}{\boldsymbol{\Lambda}}
\newcommand{\opsi}{\overline{\psi}}
\renewcommand{\tan}{\textrm{tg}}
\renewcommand{\cot}{\textrm{ctg}}
\renewcommand{\sinh}{\textrm{sh}}
\renewcommand{\cosh}{\textrm{ch}}
\renewcommand{\tanh}{\textrm{th}}
\renewcommand{\coth}{\textrm{cth}}

\begin{abstract}
 We take a null hypersurface (causal horizon) generated by a congruence of null geodesics as the boundary of the Doran - Lobo - Crawford spacetime to be the place where the Brown - York quasilocal energy is located. The components of the outer and inner stress tensors are computed and depend on time and the impact parameter $b$ of the test particle trajectory. The spacetime is a solution of Einstein's equations with an anisotropic fluid as source.
 
 The surface energy density $\sigma$ on the boundary is given by the same expression as that obtained previously for the energy stored on the Rindler horizon. For long time intervals with respect to $b$ (when the stretched horizon tends to the causal one), the components of the stress tensors become constant.

\end{abstract}


\section{Introduction}
 Einstein's equations of General Relativity, like Newton's law for gravitation, indicate that matter is the source for gravity. On the grounds of the Equivalence Principle, Khoury and Parikh \cite{KP} ask the question : if gravity depends on matter, can acceleration be atributed to matter? But the matter distribution is encoded in the stress-energy tensor. Therefore, where the energy and stresses are localized, as we could accelerate even in Minkowski space, which is empty, at least according to the actual view? 
 
 Usually, the boundary conditions for the metric are in the form of an induced metric and extrinsic curvature for some hypersurface. There has been an increasing interest in boundary matter in recent years. It appears, for example, in the so called black hole's fluid membrane model of the stretched horizon \cite{TPM} \cite{PW}. It is worth to note that the Gibbons - Hawking term in the Einstein - Hilbert action is a surface integral over the outer boundary of spacetime and not over the stretched horizon.
 
 As Khoury and Parikh have conjectured, ''matter refers to both bulk and boundary stress tensors''. They uniquely specify the geometry of spacetime. This may be recognized from the fact that the bulk stress tensor does not fully determine the Riemann tensor.
 
 Our purpose in this paper is to find the expressions for the stress tensors for the boundary matter, i.e. the Brown - York gravitational energy for the inner and outer regions of the boundary, as well as for the stress tensor constructed from the intrinsic geometry on the boundary. We start with the black hole interior geometry \cite{HC1} and take a hypersurface orthogonal to a congruence of null geodesics \cite{ND} \cite{RW} as the boundary of spacetime (it plays the same role as the stretched horizon in the Parikh-Wilczek membrane paradigm of the horizon, for long time intervals). 
 
 \section {The Doran - Lobo - Crawford spacetime}
 
 It is a well-known fact that the geometry inside the horizon of a Schwarzschild black hole is dynamic since the radial coordinate becomes timelike and the metric is time dependent. 
 It is given by 
 \begin{equation}
 ds^{2} = -\left(\frac{2m}{t}-1\right)^{-1} dt^{2} + \left(\frac{2m}{t}-1\right) dz^{2} + t^{2} d\Omega^{2} 
 \label {2.1}
 \end{equation}
 where $m$ is the mass of the black hole, $z$ plays the role of the radial coordinate, with $-\infty < z < +\infty$ and $d\Omega^{2} = d\theta^{2} + sin^{2} \theta d\phi^{2}$. Throughout the paper the velocity of light $c = 1$.

 Doran et al. \cite{DLC} have taken into consideration the case of a time-dependent $m(t)$, when the metric (2.1) is no longer a solution of the vacuum Einstein equations. A time dependent mass inside the horizon is justified by the fact that it is equivalent to  a $r$ - dependent mass $m(r)$ outside the horizon \cite {DV}, which is a widespread situation. Moreover, Lundgren, Schmekel and York, Jr. \cite{LSY} showed that the quasilocal energy of a black hole depends on the radial position of the observer outside the horizon and on time inside it . 
 
 Doran et al. observed that the spacetime acquires an instantaneous Minkowski form for $m(t) = t$. 
\begin{equation}
ds^{2} = -dt^{2} + dz^{2} + t^{2} d\Omega^{2} 
\label{2.2}
\end{equation}
 The line element (2.2) has a curvature singularity at $t = 0$ since the scalar curvature $R_{\alpha}^{\alpha} = 4/t^{2}$ is infinite there.
However,a  geodesic particle with constant angular coordinates moves exactly as in flat space. 

 In order to be a solution of Einstein's equations, we must have an energy - momentum tensor ( a source ) on the r.h.s. In the spirit of Lobo \cite{FL} and Viaggiu \cite{SV} we consider the spacetime (2.2) to be filled with an anisotropic fluid with the stress tensor 
 \begin{equation}
 T_{\mu}^{\nu} = \rho \eta_{\mu} \eta^{\nu} + p s_{\mu} s^{\nu}
 \label{2.3}
 \end{equation}
 where $\eta_{\mu} = (1,0,0,0)$ is the fluid 4 - velocity (the components are in order t, z, $\theta$, $\phi$), $s_{\mu} = (0,1,0,0)$ is the unit spacelike vector in the direction of anisotropy (with $\eta_{\alpha} \eta^{\alpha} = -1$, $s_{\alpha}s^{\alpha} = 1~~and~~s_{\alpha} \eta^{\alpha} = 0$). Hence, we have only one principal pressure $p$ and the transverse pressures measured in the orthogonal direction to $s^{\mu}$ are vanishing. 
 
 With (2.3) on the r.h.s. of the Einstein equations, one obtains, for the energy density and pressure on the $z$ - direction
 \begin{equation}
 \rho(t) = -T_{t}^{t} = \frac{1}{4 \pi t^{2}},~~~p(t) = T_{z}^{z} = -\frac{1}{4 \pi t^{2}},~~T_{\theta}^{\theta} = T_{\phi}^{\phi} = 0.
 \label{2.4}
 \end{equation}
 (The convention $R_{\alpha \beta} = \partial_{\nu} \Gamma_{\alpha \beta}^{\nu}-...$ has been used). We note that $p(t) = -\rho(t)$, as in the ''gravastars'' models with anisotropic pressures \cite{VW}).
 
 It is an easy task to check that the weak, strong and dominant energy conditions for $T_{\mu}^{\nu}$ are obeyed
 \begin{equation}
 T_{\mu \nu} \eta^{\mu}\eta^{\nu} = \rho > 0 ~~~(weak~ energy ~condition)
 \label{2.5}
 \end{equation}
 
 \begin{equation}
 ( T_{\mu \nu} - \frac{1}{2}g_{\mu \nu}T) \eta^{\mu} \eta^{\nu} = 0~~~(strong ~energy~ condition)
 \label{2.6}
 \end{equation}
 and the energy - momentum 4 - current density of the fluid \cite{RW}
 \begin{equation}
 - T_{\alpha}^{\beta} u^{\alpha} = \rho u^{\beta} 
 \label{2.7}
 \end{equation}
 is a timelike vector (dominant energy condition).
 
 Our aim is to apply the Khoury - Parikh prescription to compute the boundary stress tensor replacing the boundary conditions by boundary sources (the knowledge of the bulk geometry from Einstein's equations is not sufficient : we need also to add matter on the boundary). The authors of \cite{KP} localize the boundary stress energy on the horizon of the spacetime (be it de Sitter, Rindler or any other horizon). 
 
 The situation is different in Minkowski space : because of the lack of a horizon, the location of the boundary was chosen to be a two-sphere at some fixed large radius $r_{0}$ \cite{KP}, where the geometry is just the Einstein static universe.\\
 But what is the physical meaning of $r_{0}$ or of its order of magnitude? Our choice for the location of the boundary matter  is a null hypersurface generated by a congruence of null geodesics (the causal horizon). The null geodesics for the metric (2.2) have been computed in \cite{HC1}. From (2.2) we have
 \begin{equation}
\dot{t}^{2} - \dot{z}^{2} - t^{2} \dot{\phi}^{2} = 0 ,~~~(\theta = \pi/2)
\label{2.8}
\end{equation}
where an overdot denotes the derivative with respect to the affine parameter $\lambda$ along the null geodesic. Keeping in mind that $\dot{z} = const. \equiv p_{z}$ and $\dot{\phi} = L/t^{2}$, where the conserved quantities $p_{z}~ and~ L$ are the momentum along the z - direction per unit mass and, respectively, the angular momentum per unit mass, one obtains
\begin{equation}
z(t) = \pm \sqrt{t^{2} + b^{2}},~~~~\frac{d\phi}{dt} = \pm \frac{b}{t \sqrt{t^{2} + b^{2}}}
\label{2.9}
\end{equation}
where $b \equiv L/p_{z}$ plays the role of an impact parameter . \\We observe that the projection of the null particle trajectory on the z - axis is a hyperbola where $a ~\equiv~ 1/b$ may be interpreted as the rest-system acceleration. It is worth to note that  $d\phi/dt$ decreases very fast with time while $dz/dt$ tends to unity. Therefore we neglect $d\phi/dt$ in the regime $t >> b$, a situation when the null test particle moves approximately on the z - axis. That means to replace the causal horizon with a ''stretched'' horizon \cite{KP} and take $t \longrightarrow \infty$ in the final result for the components of the boundary (which now becomes timelike) stress tensor.

\section{The stress tensor of the boundary sources}

 Let us compute now the components of the extrinsic curvature of the timelike hypersurface $S$, given by 
\begin{equation}
z - \sqrt{t^{2} + b^{2}} = 0.
\label{3.1}
\end{equation}
Using  the Berezin- Kuzmin - Tkachev method \cite{BKT}, we obtain the components of the normal vector 
\begin{equation}
n_{\alpha} = (-\frac{t}{b}, \frac{z}{b}, 0, 0)
\label{3.2}
\end{equation}
with $z = \sqrt{t^{2} + b^{2}}~ and~ n_{\alpha} n^{\alpha} = 1$. Using (3.2), the induced metric on $S$, embedded in the spacetime (2.2), is 
\begin{equation}
h_{\alpha \beta} = g_{\alpha \beta} - n_{\alpha} n_{\beta} 
\label{3.3}
\end{equation}
Therefore, for the $tt,~ \theta \theta~ and~ \phi \phi$ components of the extrinsic curvature tensor
\begin{equation}
K_{\alpha \beta} = - \nabla_{\alpha} n_{\beta}
\label{3.4}
\end{equation}
one obtains
\begin{equation}
K_{t}^{t} = K_{\theta}^{\theta} = K_{\phi}^{\phi} = -\frac{1}{b}
\label{3.5}
\end{equation}

Let us find now the Ricci tensor on the surface (3.1), namely the intrinsic curvatures of $S$, using the Gauss - Codazzi equations \cite{MTW} \cite{CL} 
\begin{equation}
^{3}R + K_{ij} K^{ij} -(K_{i}^{i})^{2} = -2 G_{\alpha \beta} n^{\alpha} n^{\beta}
\label{3.6}
\end{equation}
and
\begin{equation}
G_{j}^{i} = ^{3}G_{j}^{i} - K_{l}^{l} K_{j}^{i} + \frac{1}{2} \delta_{j}^{i}\left[(K_{l}^{l})^{2} + K_{l}^{i} K_{j}^{l}\right] 
\label{3.7}
\end{equation}
where $i, j$ correspond to $(t, \theta, \phi)$, and $^{3}R,~ ^{3}G_{j}^{i}$ are the scalar curvature and Einstein's tensor on $S$, respectively. By means of (3.2), (3.5) and the fact that, in the spacetime (2.2) \cite{HC1},
\begin{equation}
R_{tt} = R_{zz} = R_{zt} = 0,~~~~R_{\theta \theta} = \frac{R_{\phi \phi}}{sin^{2}\theta} = 2,~~~~R_{\alpha}^{ \alpha} = \frac{4}{t^{2}}
\label{3.8}
\end{equation}
we reach at the following expressions
\begin{equation}
^{3}R = \frac{4}{t^{2}} + \frac{6}{b^{2}},~~~^{3}R_{t}^{t} = 0,~~~^{3}R_{\theta}^{\theta} = ^{3}R_{\phi}^{\phi} = \frac{2}{t^{2}} + \frac{3}{b^{2}}
\label{3.9}
\end{equation}
and
\begin{equation}
^{3}G_{t}^{t} = -\frac{2}{t^{2}}-\frac{3}{b^{2}},~~~^{3}G_{\theta}^{\theta} = ^{3}G_{\phi}^{\phi} = 0.
\label{3.10}
\end{equation}

We are now in a position to find the expressions of the quasi-local stress energy $T_{\alpha \beta}^{in}$ defined on the boundary of the inner spacetime region $M_{in}$ 
\begin{equation}
8 \pi G_{4}~ T_{\alpha \beta}^{in} = K_{\nu}^{\nu}~ h_{\alpha \beta} - K_{\alpha \beta}
\label{3.11}
\end{equation}
and of the stress tensor $T_{\alpha \beta}^{out}$ for the boundary matter (the Brown - York holographic stress tensor for the outer spacetime $M_{out}$) \cite{KP} 
\begin{equation}
^{3}G_{\alpha \beta} = 8 \pi G_{3} ~( T_{\alpha \beta}^{in} + T_{\alpha \beta}^{out})
\label{3.12}
\end{equation}
From the viewpoint of an observer in $M_{in}$, $S$ encodes the effects of $M_{out}$. In the previous equations, $G_{4}~ and~ G_{3}$ are the 4-dimensional Newton's constant and the 3-dimensional one, respectively, $^{3}G_{\alpha \beta}$ is the intrinsic Einstein tensor on the boundary surface, obtained from (3.10).

The equations (3.11) and (3.5) yield
\begin{equation}
8 \pi G_{4}~ T_{t}^{t,in} = 8 \pi G_{4}~ T_{\theta}^{\theta, in} = 8 \pi G_{4}~ T_{\phi}^{\phi,in} = -\frac{2}{b}
\label{3.13}
\end{equation}
 We see that $T_{\mu}^{\nu,in}$ is time independent. $\rho (t)$ and $p(t)$ from (2.4) tends to zero when $t \longrightarrow \infty$ and, therefore, in that case there is a contribution only from the boundary to the stress tensor measured by an interior observer.
Combining (3.10) and (3.13), one obtains
\begin{equation}
-\frac{2 ~G_{3}}{b~ G_{4}} + 8 \pi G_{3}~\textsl{\textit{}} T_{t}^{t,out} = -\frac{2}{t^{2}}-\frac{3}{b^{2}}
\label{3.14}
\end{equation}
and
\begin{equation}
-\frac{2 ~G_{3}}{b~ G_{4}} + 8 \pi G_{3}~ T_{\theta}^{\theta,out} = 0
\label{3.15}
\end{equation}

$G_{3}$ has dimension $G_{4}/distance$. Keeping in mind that we have only one constant with dimension of length at our disposal - the impact parameter $b = L/p_{z}$, we conjecture the two Newton's constants $G_{3}$ and $G_{4}$ are related by  
\begin{equation}
\frac{G_{3}}{G_{4}} = \frac{1}{b}, 
\label{3.16}
\end{equation}
 as in the brane-world scenarios. The parameter $b$ plays the same role as $r_{0}$ for the Minkowski space in the Khoury - Parikh model. Therefore, the components of $T_{\alpha \beta}^{out}$ are given by 
\begin{equation}
T_{t}^{t,out} = -\frac{1}{8 \pi G_{3}} (\frac{2}{t^{2}} + \frac{1}{b^{2}})
\label{3.17}
\end{equation}
and 
\begin{equation}
T_{\theta}^{\theta,out} = T_{\phi}^{\phi,out} = \frac{1}{8 \pi G_{3}} \frac{2}{b^{2}}
\label{3.18}
\end{equation}

Let us notice that, in our sign conventions, $T_{t}^{t,in}$ is minus the surface energy density $\sigma$ on $S$. Keeping in mind that $b = 1/a$, we have
\begin{equation}
a = 4 \pi G_{4} \sigma ,
\label{3.19}
\end{equation}
a result already obtained in \cite{HC2} for the energy stored on the Rindler horizon. The result is not surprising. From eq. (3.11) we see that $T_{t}^{t,in}$ is determined by the extrinsic curvature and the metric on the boundary of $M_{in}$. Its physical meaning is the surface energy density on the boundary, viewed from $M_{in}$. The projection (2.9) of the null particle trajectory on the z - axis is a hyperbola, with the proper acceleration $a = 1/b$, as for a Rindler observer . When $t >> b$, the surface $S$ becomes null (the causal horizon) since $z(t) \approx t$ and the term $2/t^{2}$ may be neglected  from the expression of $T_{\alpha \beta}^{out}, T_{\alpha \beta}^{in}$ and all the components depend only on the impact parameter $b$. From (3.16) one infers that a very large $b$ leads to the cancellation of gravity ($G_{3}$ tends to zero) on the boundary $S$.

 We finally mention that our surface stress energy  computations are in accordance with Mach's Principle on the grounds of the fact that the gravitational field stress tensor is localized on the boundary of the region considered, namely the hypersurface $z(t)$ from (2.9) (the ''stretched'' horizon - a timelike surface hovering just inside the causal horizon \cite{KP})  which, when $t$ tends to infinity, becomes the event horizon. In other words, boundary conditions are replaced by boundary sources which play the same role as ''the distant stars'' .

\section{Conclusions}

On the same line as Khoury and Parikh, we underlined in this paper the role played by the boundary matter in determining the geometry of a spacetime, besides the bulk matter, acquiring in that way a Machian character.

The boundary where the Brown - York gravitational energy is localized was chosen to be the causal horizon of the Doran - Lobo - Crawford geometry, generated by a congruence of null geodesics.

Using the $3+1$ ADM splitting of Einstein's equations, we found the intrinsic geometry on the ''stretched'' horizon and the components of the interior and exterior stress tensors corresponding to the Brown - York gravitational energy of the two spacetime regions.

\end{document}